\documentclass[unnumsec,webpdf,modern,large]{mam-authoring-template}%

\usepackage[version=4]{mhchem}

\graphicspath{{figures/}}

\theoremstyle{thmstyleone}%
\theoremstyle{thmstyletwo}%
\theoremstyle{thmstylethree}%

\begin{document}

\journaltitle{Microscopy and Microanalysis}
\DOI{DOI HERE}
\copyrightyear{2026}
\pubyear{2026}
\access{Advance Access Publication Date: Day Month Year}
\appnotes{Original Article}

\firstpage{1}


\title[Using superpixels for interpretable feature reduction in large 2D diffraction datasets]{Using superpixels for interpretable feature reduction in large 2D diffraction datasets}

\author[1]{Andreas Werbrouck\ORCID{0000-0003-3796-0024}} 
\author[2]{Nikhila C. Paranamana$\ast$\ORCID{0000-0001-9466-5780}}
\author[3]{Andrew C. Meng$\ast$\ORCID{0000-0002-3060-8928}}
\author[4,5]{Xiaoqing He$\ast$\ORCID{0000-0001-6058-9732}}
\author[1,2,6]{Matthias J. Young$\ast$\ORCID{0000-0001-7384-4333}}

\authormark{Werbrouck et al.}

\address[1]{\orgdiv{Materials Science and Engineering Institute}, \orgname{University of Missouri},  \orgaddress{416 S 6th St \postcode{65201}, \state{Columbia, MO}, \country{USA}}}
\address[2]{\orgdiv{Department of Chemistry}, \orgname{University of Missouri}, \orgaddress{601 S College Ave \postcode{65201}, \state{Columbia, MO}, \country{USA}}}
\address[3]{\orgdiv{Department of Physics}, \orgname{University of Missouri}, \orgaddress{701 S College Ave, \postcode{65201}, \state{Columbia, MO}, \country{USA}}}
\address[4]{\orgdiv{Electron Microscopy Core}, \orgname{University of Missouri}, \orgaddress{1030 Hitt St. \postcode{65201}, \state{Columbia, MO}, \country{USA}}}
\address[5]{\orgdiv{Institute of Materials Science and Engineering}, \orgname{Washington University in Saint Louis}, \orgaddress{1 Brookings Drive, \postcode{63130} \state{St. Louis, MO}, \country{USA}}}
\address[6]{\orgdiv{Department of Chemical and Biomedical Engineering}, \orgname{University of Missouri}, \orgaddress{416 S 6th St  \postcode{65201}, \state{Columbia, MO}, \country{USA}}}

\corresp[$\ast$]{Corresponding author. \href{email:matthias.young@missouri.edu}{matthias.young@missouri.edu}}

\received{Date}{0}{Year}
\revised{Date}{0}{Year}
\accepted{Date}{0}{Year}

\abstract{Large 2D diffraction datasets, consisting of hundreds or thousands of measurements, are commonly acquired in scanned nanobeam diffraction, or through scanning transmission electron microscopy (STEM) with pixelated detectors, or at synchrotron X-ray beamlines. Machine learning and artificial intelligence offer great promise for analyzing these datasets. However, the sheer volume of data presents a significant processing bottleneck. Cropping the detector and pixel binning are standard ways to reduce data size. Here we propose grouping and averaging pixels into superpixels of variable area. High-information regions are sampled densely, while low-information areas are collected into larger superpixels. In the process, symmetries in the data are captured and exploited, making this approach suitable for preprocessing 2D diffraction data. We compare two variance-minimizing methods: K-means clustering (top-down) and agglomerative clustering (bottom-up) and demonstrate superior scaling and interpretability for the bottom-up method. As these methods are distance-based, we demonstrate that the construction of superpixels can be accelerated using Gaussian random projection. Finally we show over 100-fold acceleration for phase mapping with non-negative matrix factorization on a 4D-STEM dataset when superpixels are used as a preprocessing step.}

\keywords{Scanning transmission electron microscopy, Scanned nanobeam diffraction, 4D-STEM, Matrix decomposition, Superpixels}

\maketitle

\section{Introduction}
\subsection{Context}
Diffraction is a key tool for structural analysis in various domains of science. It offers direct access to reciprocal space, which can be used to probe characteristic length scales of a wide range of structures, from atomic structures of materials \citep{Krawitz2001} to structural biology \citep{Shi2014}. 

In the case of macroscopic (powder) diffraction any preferred directions are averaged out, so a 1D (or scanned 0D/point) detector is suitable to collect the available data. However, when structures with preferred orientation are considered, or small volumes are probed with a fine beam, this averaging over all angles does not occur. In such cases, it is beneficial to probe the reciprocal space using 2D area detectors. Typical examples of such situations are electron diffraction in a transmission electron microscope (TEM) \citep{Bendersky2001}, or X-ray diffraction experiments at dedicated synchrotron beamlines (e.g. \citep{Bernasconi2014, Liermann2015, Allan2015, Dippel2015, Schkel2021, Fitch2023, Kawaguchi2024, Zimmermann2025})

\begin{figure}[!ht]
\centering
\includegraphics[width=\columnwidth]{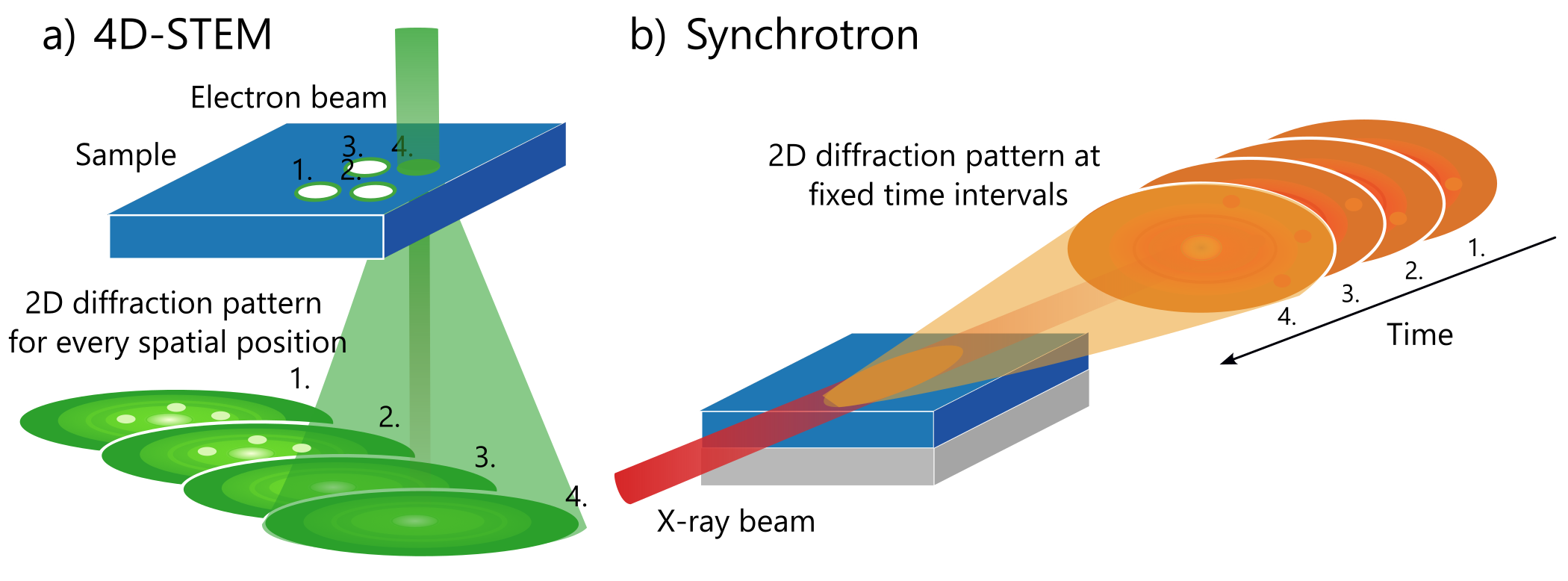}
\caption{\label{fig_1} a) In 4D-STEM or b) during a synchrotron measurement, hundreds of diffractograms may be collected, either separated in space or in time. The interpretation of these datasets to the fullest extent is often challenging due to their large size.}
\end{figure}

More powerful diffraction measurements have been enabled by the development of fast, radiation-hard 2D detectors with high dynamic range and noise-free detection capabilities for electrons, based on techniques from particle physics \citep{MacLaren2020}. 

Examples for detecting electrons are Electron microscopy pixel array detector (EMPAD) \citep{Tate2016High}, MerlinEM \citep{Krajnak2023}, and Dectris Arina based on the KITE ASIC \citep{Zambon2023}. This has lead to strong developments in the field of Four-Dimensional Scanning Transmission Electron Microscopy (4D-STEM) \citep{Nord2020, Paterson2020}.

Similarly, for X-rays, examples of 2D detectors are PILATUS \citep{Broennimann2006PILATUS}, EIGER \citep{Dinapoli2011EIGER}, or the (custom-built) CirPAD \citep{Desjardins2022}. These detectors, combined with bright synchrotron sources, offer fast acquisition rates, large pixel counts and high dynamic ranges, leading to the proliferation of \textit{in situ} diffraction experiments. 

4D-STEM (Figure \ref{fig_1}a) \citep{Carter2016Transmission} is an electron microscopy technique in which a very fine electron beam is scanned over small or large areas ranging from 10$^0$-10$^6$ nm$^2$, depending on the application. As the beam scans and is transmitted through the sample, its electrons scatter off the atoms in the material and can be collected as diffraction patterns. Hence, a 4D-STEM dataset consists of a collection of 2D diffraction patterns, one for every spatial pixel \citep{Ophus2019Four}. Depending on the beam convergence angle, the internal structure of the central beam or diffraction peaks can reveal additional information about the sample, adding to the appeal of collecting 4D data \citep{Wu2023}.

Similarly, high-brightness electromagnetic radiation is available at synchrotron facilities to characterize all kinds of matter \citep{Sokolov1968Synchrotron}, and specifically X-ray diffraction has been a workhorse in this respect \citep{He2018Two}. Some typical \textit{in situ} experiments, relevant in the context of this work, include collecting 2D X-ray diffraction data while annealing a material \citep{Sawayanagi2006}, performing electrochemical charge-discharge cycles \citep{Young2017}, inducing ferroelectric switching \citep{Khamidy2017}, or even in situ synthesis of materials \citep{Young2020, Saha2015} (Figure \ref{fig_1}b). 

Such 2D diffraction datasets can be very large, with sizes exceeding several gigabytes or more \citep{Wang2018Synchrotron} \citep{Li2021Machine}. For example, a $2048 \times 2048$ diffractogram with a bit depth of 32 takes about 16 MB of memory to store. If on a synchrotron beamline a spectrum is collected every 12 s, which is not uncommon, an annealing experiment of 3h would yield a 14.4 GB dataset. Similarly, in 4D-STEM, a $512  \times 512$, 32-bit diffractogram can be collected over $100  \times  100$ spatial positions in a matter of minutes, or even seconds depending on the detector, yielding a dataset of 10.4 GB. 

\subsection{Related work: analysis approaches for large diffraction datasets} 
Because of their large size and dimensionality, 4D-STEM and synchrotron datasets offer a rich testbed for computer-assisted interpretation and visualization. Although there are some idiomatic differences between spatial mapping and timeseries collection, ultimately, the desired end result in both cases is often the same: a map (2D) or plot (1D) of one or more properties of interest (phase, orientation, strain, thickness, etc.) as a function of space or time. 
Hence, in both instances a typical analysis pipeline will work alternatingly on reciprocal-space pixels and real-space pixels to reduce dimensionality through several steps of clustering, projecting, integrating and factorization. 

One method that is useful in this workflow is component analysis, in which e.g. principal component analysis (PCA), nonnegative matrix factorization (NMF), or related techniques are used to determine a (usually small) number of physically interpretable components: fingerprints in reciprocal space, and maps in real space (or timeseries). NMF has been used to analyze combinatorial XRD libraries (1D patterns) \citep{Long2009, Stanev2018}, and \cite{Kutsukake2024} used NMF on a series of 2D synchrotron measurements on SiGe alloy. Also in the 4D-STEM field,  interest in NMF is growing \citep{Eggeman2015, Martineau2019, Uesugi2021Non, Allen2021fast, Thronsen2024, Kang2024Mapping, Kimoto2024Unsupervised, Yoo2024, kimoto2025nonnegative}. 
In some instances NMF has been employed on features (Bragg discs) instead of full patterns \citep{Bruefach2022}. \citep{Kimoto2024Unsupervised} and \citep{Cao2025} performed additional clustering on the components obtained through NMF. Our recent work focused on enabling faster NMF directly on large 4D-STEM measurements \citep{Werbrouck2025Fast} through QB factorization as a preprocessing step (Randomized NMF or RNMF). In a way, the hierarchical alternating least squares (HALS) algorithm in NMF is an extreme embodiment of the alternating approach described above, which proceeds by gradually improving projections along both real-space and reciprocal axes. 

However, NMF is also non-convex and can be time-consuming. As an alternative, autoencoders can be used. Autoencoders do not provide easily interpretable fingerprints (neither does PCA) but autoencoders are usually better suited for non-linear data than PCA. Recent examples include \citep{Bridger2023, Ludacka2024, Kho2025}, which mapped latent features of an autoencoder to construct spatial maps from 4D-STEM data, and \citep{Strohmann2023}, which mapped the latent variable of a variational autoencoder to distinct phases in a large, time-dependent HEXRD dataset. 

Beyond component analysis, direct clustering in real space is another computationally efficient strategy for data reduction. The simplest approach is to cluster real space pixels with K-means, using the reciprocal pixels as features \citep{Ren2025}. Recently, two works presented novel ways to group 4D-STEM pixels in real space: \citep{Lee2026} use a marching squares algorithm to group real-space pixels into clusters, averaging diffraction signatures in the process. \citep{Shi2022} and \citep{MacLaren2026} use K-means and DBSCAN, respectively, to group real-space pixels using bragg peaks as features as opposed to the raw diffraction data. 

The identification of structures from the diffraction patterns is a distinct problem from clustering. Structure identification can be performed through template matching, or automated crystal orientation matching (ACOM) after automated disc detection \citep{Cautaerts2022, ophus2022automated, Jeong2021}. Finally, deep learning has also been proposed for structure identification \citep{Ra2021, Chen2023, Mika2024, Gleason2024, Nathani2026} and stress mapping \cite{Yuan2021, Munshi2022}. 

\subsection{Problem identification}
The choice of data analysis approach ultimately depends on what one wants to extract from the experiment. However, due to the high detector resolutions, computational demands for conducting any of the aforementioned analyses can escalate quickly, beyond what is tractable with current technology \citep{Werbrouck2025Fast}. Pixel binning and/or detector cropping are default solutions to this problem, leading to smaller, more manageable datasets. In reciprocal space, summed or averaged binning has the additional advantage that decreased spatial resolution is traded for a higher dynamic range. Because binning and cropping are so intuitive, they are hardly perceived as feature reduction strategies. Peak detection may work well for crystalline materials but requires a detection threshold and is less suited for amorphous and mixed materials. 

In this work, we demonstrate that grouping reciprocal pixels into \textit{superpixels} can be a very powerful feature reduction technique. The pixels in the 2D detector array are clustered into a set of superpixels, which remain the same throughout the entire dataset. These superpixels then can serve as a drop-in featurization before any other analyses of choice (factorization, real-space clustering, etc... ) is performed.

In the first part, we explore different methods of creating superpixels, evaluating them for faithful reconstruction of the original dataset, computational efficiency, and interpretability. Then, we focus on accelerating the construction of superpixels. Finally, we show that the resulting feature reduction, depending on the chosen algorithm and number of superpixels, provides a 100-fold improvement in the speed with which phase mapping can be performed in a 4D-STEM dataset without sacrificing accuracy.

\subsection{Proposed strategy}
In diffraction patterns, there is often symmetry around the beam center (with specific spots and/or rings present depending on the phase). As such, the data in many pixels is redundant and the information density is different in different areas of the detector. At lower scattering angles (closer to the beam center) or at diffraction spot positions, the information density is higher, whereas more pixels carry less information at higher scattering angles. Most importantly, the locations of high and low information density on the detector generally stay roughly the same during a series of measurements. Ideally, one would want to densely sample regions where information density could be high (small bins around the center and anywhere diffraction peaks occur at some point during the measurement), while more pixels could be aggregated in low-count regions (larger bins in between peaks). 

Consider a diffraction dataset stored in a 3D matrix $X_{mn_xn_y}$ consisting of $m$ diffraction measurements over a 2D detector grid of $n_x$ horizontal pixels and $n_y$ vertical pixels. For the purpose of identifying detector regions which can be aggregated, this 3D matrix can be reshaped into a 2D matrix $X_{mn}$ for analysis. Each 2D diffraction pattern (of size $n_x \times n_y$) is unrolled into a row of length $n = n_x \times n_y$. 

We propose that it is possible to exploit the properties of diffraction to decrease the length of the rows in $X_{mn}$  from $n$ detector pixels to $k$ superpixels with $k \ll n$ to obtain $X_{mk}$ for further processing.

When grouping and averaging pixels with the aim of later reconstructing a 2D signal, it is important to create clusters such that their values are as close as possible to the superpixel average. In other words, the objective is to minimize the sum of variances for all pixels $x$ within each superpixel $S_i$. That is, we want find a set of superpixels $S_1,\ldots,S_k$ such that

\begin{equation}
    f\left(S\right)=\sum_{i=1}^{k}\sum_{x\ \in\ S_i}\left|\left|x\ -\ \mu_i\right|\right|^2 
    \label{eq_objective}
\end{equation}

is minimal. $x$ and $\mu_i$ are vectors with dimension $m$ corresponding to the number of diffraction patterns collected. Furthermore, the number of pixels $a_i$ in superpixel $S_i$ is not equal for all superpixels.

We consider two approaches to achieve this minimization: a top-down (vector quantization) and a bottom-up (agglomerative clustering) method. Each of these has distinct characteristics and use cases when used on images \citep{Felzenszwalb2004Efficient, Achanta2012SLIC}. Both also find use cases beyond image analysis. In vector quantization all pixels start in a large cluster, which is then further divided using a K-means approach to minimize the objective function $f$ (Eq. \ref{eq_objective}) in a top-down fashion. The ensemble of centroids $\mu_i$ is known as a codebook. In agglomerative clustering, every pixel initiates in its own cluster. Clusters are then merged according to specific rules (bottom up). Similarly to vector quantization, these rules can be chosen such that clusters are merged to minimize $f$ (again, Eq. \ref{eq_objective}) and thus the reconstruction error (i.e. Ward's method). An optional additional constraint is to only consider the local neighborhood for merging candidates. 

\begin{figure*}[!t]
\centering
\includegraphics[width=\textwidth]{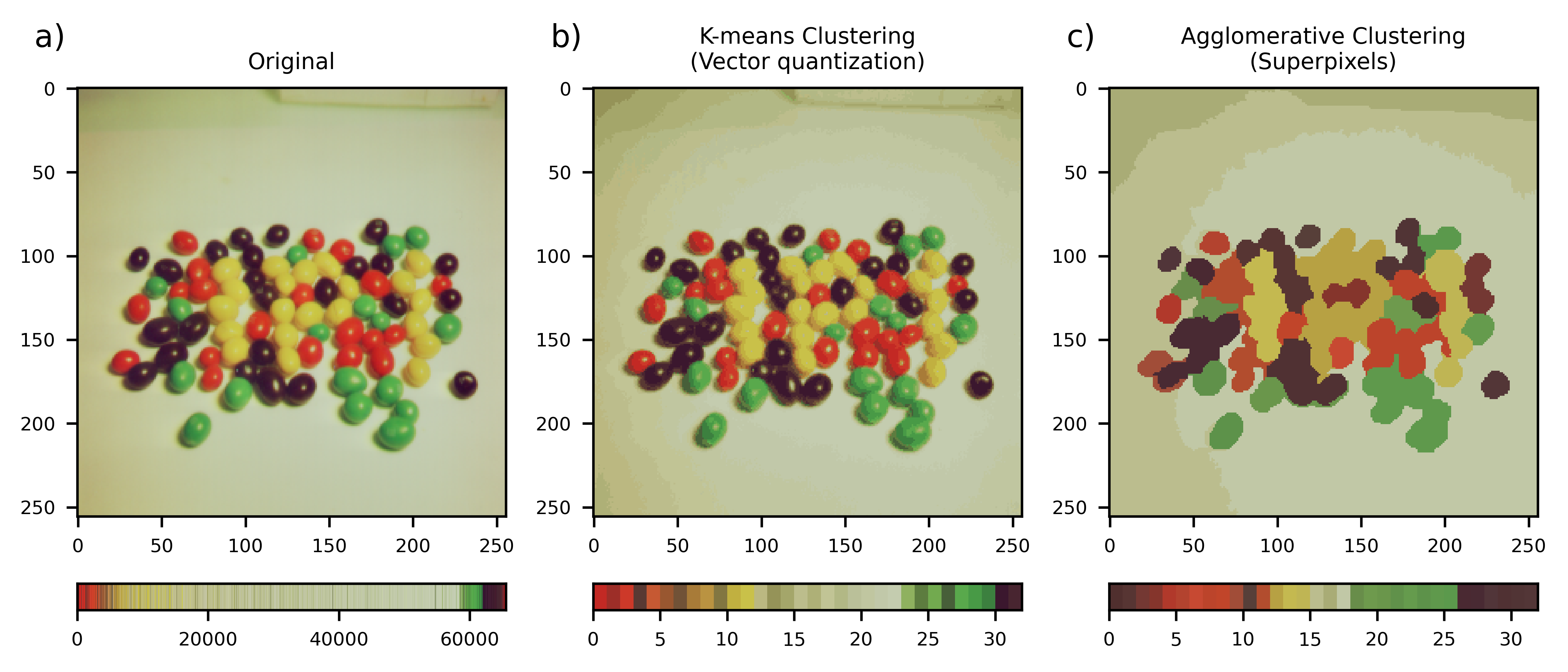}%
\caption{\label{fig_2} Vector quantization compared to agglomeration approaches for clustering pixels in an image with $m=3$ channels: red, green and blue. a) original image b) pixel averaging using vector quantization c) pixel averaging using agglomerative clustering. While vector quantization with K-means clustering identifies average `colors', agglomerative clustering identifies groups of adjacent pixels as objects. The relevance of these two approaches for 2D diffraction images is discussed in the text. The bottom colorbars show the different colors present in the respective pictures.}
\end{figure*}

To illustrate these concepts, Figure \ref{fig_2} shows an example of vector quantization and superpixel construction on a real space test image. The original image (Figure \ref{fig_2}a) contains more than $60,000$ unique 24-bit colors (features). Here, by analogy, $m=3$ (red, green and blue channels being different `measurements') and $n=256\times256=65,536$. 

Using K-means-based vector quantization (Figure \ref{fig_2}b) a codebook (set of centroids) can be constructed which comprises $k=32$ colors and captures most of the color variation in the figure, reducing the number of unique colors in the image to 32 (or any arbitrary number). The color of each pixel is then replaced with the closest of the 32 codebook values. There is no emphasis on spatial distribution of these centroids because the pixels are not connected.

In Figure \ref{fig_2}c, we have used agglomerative clustering with spatial connectivity to construct 32 spatially connected superpixels. In this method, each pixel starts in its own cluster, and at each step the two most similar adjacent clusters are merged.

In both cases, we start from a matrix $X_{mn}$ with $m = 3$ and $n = 65536$, which we reduced to $X_{mk}$ with $k = 32$. Then further processing can be performed on $X_{mk}$, after which the 32 centroids $\mu_i$ can be mapped to the $256 \times 256$ positions. The value of 32 is intentionally low to exaggerate the effect of clustering on the image in Figure \ref{fig_2}. Nevertheless, this analysis highlights how clustering can be used to reduce the feature space for a given image, with specific benefits for top-down or bottom-up approaches.

While both vector quantization and agglomerative clustering group redundant pixels according to their information value, they obtain distinct results in real space: vector quantization yields an average color space, while agglomerative clustering identifies objects. Here, we posit that in reciprocal space, both methods group pixels with similar information. 

Considering this, a key difference between the example presented in Figure \ref{fig_2} and large diffraction datasets is the $m$ dimension: in the case of a color photograph (e.g. Figure \ref{fig_2}), the $m$ vector is small ($m = 3$) and unordered, while in the case of a diffraction dataset, $m$ can be very large ($m \gg 3$) and can possess spatial or temporal order. In this work, this spatial/temporal order is not taken into account beyond the local neighborhood in reciprocal space for agglomerative clustering. However, we show that even this simple consideration of local neighborhood is sufficient to provide significant advantages in explainable feature reduction.

A second difference between the example presented in Figure \ref{fig_2} and large diffraction datasets is that the naive clustering approach presented in Figure \ref{fig_2} breaks down when a series of images (e.g. video) are considered ($m \gg 3$). Lighting may change, altering the color space. Or an object of interest may move in the field of view. In other words, it may be necessary to frequently reconsider the codebook/superpixels. However, in diffraction images, the symmetry in a collection of diffraction images from the same material in the same detection environment, with the same sample and detector position, etc. allows us to use the same set of superpixels on an entire dataset.

\section{Methods}
\subsection{Mean square error calculation}
To define error metrics, we can rewrite eq. \ref{eq_objective} as finding S such that

\begin{equation}
    f\left(S\right)=\sum_{j=1}^{m}\sum_{i=1}^{n}\left(x_{ij}\ -\ \mu_{ij}\right)^2 
    \label{eq_objective_rewritten}
\end{equation}

is minimal. Here the first summation captures the dimension of every pixel, and the two summations in equation \ref{eq_objective} are encompassed in the second summation. Similarly, we can weigh f to obtain the (scalar) mean square error for the optimal set $S$ as

\begin{equation}
    \epsilon=\frac{1}{mn}\sum_{j=1}^{m}\sum_{i=1}^{n}\left(x_{ij}\ -\ \mu_{ij}\right)^2 
    \label{eq_error_all}
\end{equation}

where $\mu_{ij}$ is the cluster center of the superpixel $S$ to which pixel $x_{ij}$ belongs. Since $m$ refers to a number of spatial or temporal diffraction measurements, in the case of spatially resolved measurements, we can discern $m_x$ and $m_y$, similar to how we defined $n_x$ and $n_y$. 

It is challenging to visualize errors in 3 or 4 dimensions, so we can vary the strategy for averaging error. If we want to map the average reconstruction error $\epsilon_n$ we can calculate 

\begin{equation}
    \epsilon_{j (0\rightarrow m)} = \epsilon_m = \frac{1}{n}\sum_{i=1}^{n}\left(x_{ij}\ -\ \mu_{ij}\right)^2 
    \label{eq_error_real}
\end{equation}

This yields an $m$-dimensional error, such that we can plot a temporal evolution of the reconstruction error, or (here) maps of the spatial regions with the largest errors.

Similarly, we can calculate

\begin{equation}
    \epsilon_{i (0\rightarrow n)} = \epsilon_n = \frac{1}{m}\sum_{j=1}^{m}\left(x_{ij}\ -\ \mu_{ij}\right)^2 
    \label{eq_error_recip}
\end{equation}

to end up with a map in the detector (reciprocal) space.

In summary, all reported reconstruction errors are calculations of mean square error (MSE), with dimensionality and averaging axes dictated by the context.

For non-negative matrix factorization in the final section (see \citep{Werbrouck2025Fast}), we approximate the input data $X \approx WH$. $X$ is an $m\times n$ matrix, and W and H $m \times l$ and $l\times n$ matrices, respectively (where $l$ signifies the number of components). We define the reconstruction error as 

\begin{equation}
    \epsilon_{NMF} = \frac{1}{mn}\sum_{i=1}^{n}\sum_{j=1}^{m}\left(x_{ij}\ -\ (WH)_{ij}\right)^2 
    \label{eq_error_nmf}
\end{equation}

where again we average over different axes for visualization in figure S3. 

In the case where we combine superpixels and (R)NMF, $X$ is transformed from an $m \times n$ matrix to an $m\times k$ matrix using superpixels, reducing the size of the problem before RNMF is employed. Separate errors can then be calculated and compared for each step in the processing pipeline using eq. \ref{eq_objective_rewritten} and \ref{eq_error_nmf}.

\subsection{Datasets}
In what follows, we will focus on 4D-STEM diffraction datasets. Every dataset was normalized by dividing it by its global maximum before processing. No baseline values were subtracted. 

The first dataset is a mixed crystalline/amorphous dataset collected in \citep{ParanamanaUnderstanding}. Briefly, this dataset was collected at the interface between LiNi$_{0.6}$Co$_{0.2}$Mn$_{0.2}$O$_{2}$ (NMC) and Li$_{10}$GeP$_{2}$S$_{12}$ (LGPS) and was collected by rastering an electron beam over a 315 nm $\times$ 180 nm  real space area (63 $\times$ 36 pixels), with a 2D diffraction spectrum collected at each measurement pixel, corresponding to 2268 total measurements. This dataset provides a good example of a real-world dataset which includes experimental uncertainty, different crystalline and amorphous phases, and variable regions of interest in the diffraction data. In this first section, we selected 16 random spatial points from this dataset for illustrative purposes. The number of diffractograms was intentionally chosen to be small, both for computational convenience and to be able to present all used data in the supporting information (Figure S1). The detector data for these 16 points was cropped into a 128$\times$128 square centered around the beam. The square was then flattened, to form a 16384 (reciprocal detector pixels) $\times$ 16 (real space positions) dataset. The connectivity among the 16384 pixels was defined based on the unflattened data. 

The second dataset is a 4D-STEM dataset collected by Ren \textit{et al.} \citep{Ren2025}, consisting of (32 $\times$ 32) real pixels $\times$ (512 $\times$ 512) diffraction pixels, cropped to 256 $\times$ 256 diffraction pixels before analysis. In the original work, it was demonstrated through various TEM-based methods that upon straining, both ferroelectric and chiral phase separation occur in \ce{BaTiS3}, and that the formation of these phases is coupled. Here this dataset is repurposed to demonstrate how the use of superpixels can accelerate phase mapping.

\subsection{Software}
Python and scikit-learn \citep{Pedregosa2011Scikit} implementations of the methods were used for all analyses. 
For K-means, k++ initialization was used. For agglomerative clustering, four-neighbor grids were used. For random Gaussian projection, an empirical value of 32 for the target dimension was used. 

A laptop with 16-core AMD Ryzen 9 8945HS and 32 GB of RAM was used for all experiments.

\section{Results}
\subsection{Superpixel construction}
\begin{figure}[!ht]
\centering 
\includegraphics[width=\columnwidth]{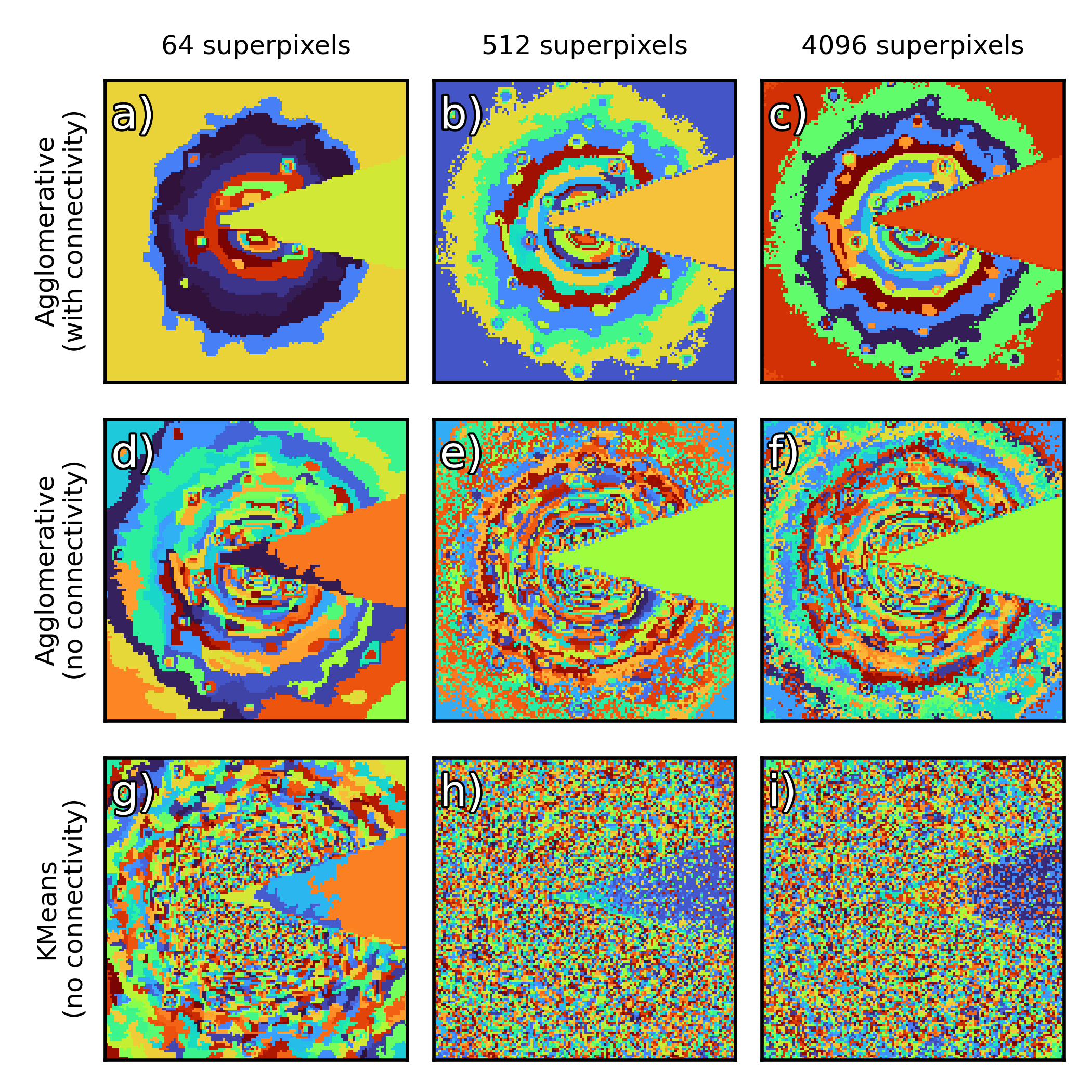}
\caption{Visualization of pixel groups with 64, 512 and 4096 superpixels, based on a limited dataset of 16 images, 3 of which are displayed in Figure \ref{fig_4}. All frames are displayed in Figure S1. a-c) Agglomerative clustering with connectivity/local structure leads to semi-connected groups of pixels. d-f) When no local structure is passed to the agglomerative clustering algorithm, clusters are less connected, but underlying symmetry is still captured in the shape. g-i) K-means clustering results in disconnected clusters.}
\label{fig_3}
\end{figure}

We investigate both top-down and bottom-up distance-based methods for pixel clustering. The reconstruction error, execution time, and interpretability are evaluated, and the discussion examines why other methods may be less suitable. 
In Figure \ref{fig_3} we depict an example of how vector quantization and superpixel methods featurize 2D diffraction data. $n=16$ spectra were randomly selected from the original dataset, cropped into $128\times128$ diffractograms around the beam center and used as a demonstration dataset with $k \in$ 64, 512 and 4096 superpixels with (variable) area $a_k$, such that the dataset of size $m \times n = 16 \times 16384$ can be compressed into a dataset with size $16 \times k$. 
The top row of Figure\ref{fig_3}  (Figure \ref{fig_3} a-c) shows agglomerative clustering with connectivity, resulting in confined superpixels. This is highly interpretable. High-information zones such as diffraction peaks are sampled densely, whereas zones far from the center (which receive less incident electron counts) are covered with larger pixels. It is meaningful that the triangular beamstop is aggregated into a single pixel: since there would be no interesting signal from this part of the detector, these pixels can all be safely averaged without losing information. 

When the connectivity constraint is removed from the agglomerative clustering (Figure \ref{fig_3} d-f), superpixels become more spatially scattered as clusters can merge with similar clusters without being spatially connected. This also results in a higher superpixel density farther away from the beam center. Even without biasing the agglomeration of neighboring pixels, the clusters form visible circular groupings, which indicate the expected intrinsic rotational symmetries in the data. 

Finally, the K-means approach, despite minimizing the reconstruction error, visually groups pixels of random spatial positions. This does not impair performance, but hampers the interpretation of the grouping (Figure \ref{fig_3} g-i). The patchiness signifies how similar neighboring pixels are, and is especially noticeable when only 64 superpixels are used. However, although the patches are discernable within the 64 superpixels, there are many more than 64 patches, and the superpixels created using the K-means approach are disjointed except for some regions of the beamstop. 
It is worth pointing out again that each set of superpixels displayed in Figure \ref{fig_3} is used to compress all 16 frames in the same way. For example, the smaller clusters localized in regions containing diffraction peaks in some frames will also be sampled more densely even if these diffraction peaks are not present in a given frame.  

\begin{figure}[!ht]
\centering
\includegraphics[width=\columnwidth]{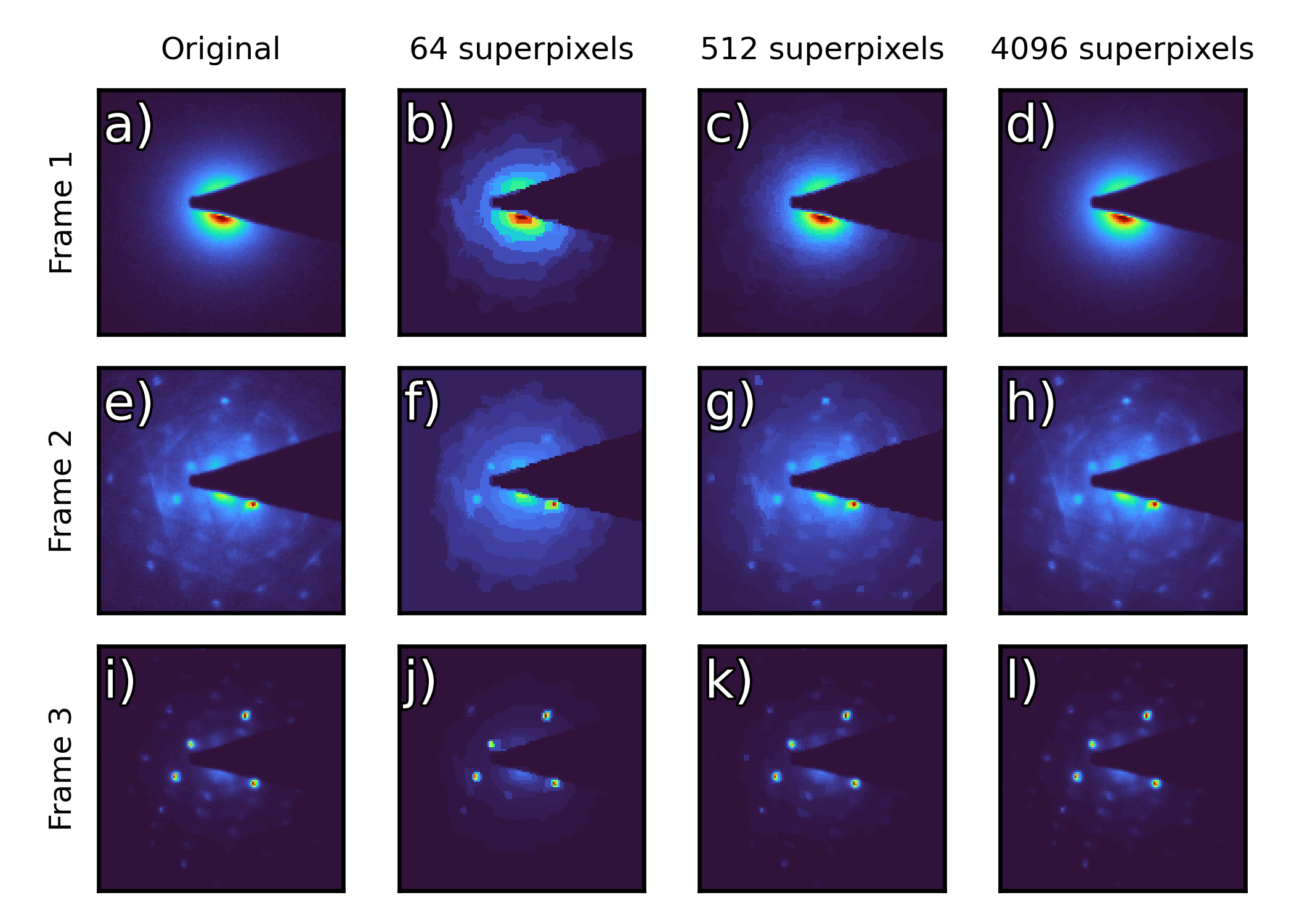}
\caption{\label{fig_4} Reconstruction of 3 frames with different numbers of superpixels, as mapped in Figure \ref{fig_3} a-c.  Note the similar peak positions in frame 2/3, despite the otherwise large differences. }
\end{figure}

As agglomerative clustering with connectivity seems to be the most interpretable pixel grouping strategy, we showcase a reconstruction of 3 frames from the dataset with these superpixels in Figure \ref{fig_4}. The superpixel mappings are the same as in Figure \ref{fig_3}a-c. The reconstruction error decreases as $k$ increases. Using 4096 clusters, it is possible to reconstruct the 3 frames (and the frames that are not displayed) very well, while scaling down the problem by a factor of 4. Inherently, this way of grouping pixels preserves more information than $2\times2$ square binning approaches that are frequently used for ad-hoc feature reduction. 

Figure \ref{fig_3} and Figure \ref{fig_4} serve to illustrate the concept of employing superpixels for feature reduction with a very limited dataset. To fully evaluate and appreciate different algorithms, we now further increase the data space to approach real-world datasets. For contrast and brevity, we abandon the agglomerative clustering without connectivity in what follows, and all instances of agglomerative clustering are performed with connectivity constraints. We also introduce random projection prior to agglomerative clustering to accelerate data processing, as discussed further below.

\begin{figure}[!ht]
\centering
\includegraphics[width=\columnwidth]{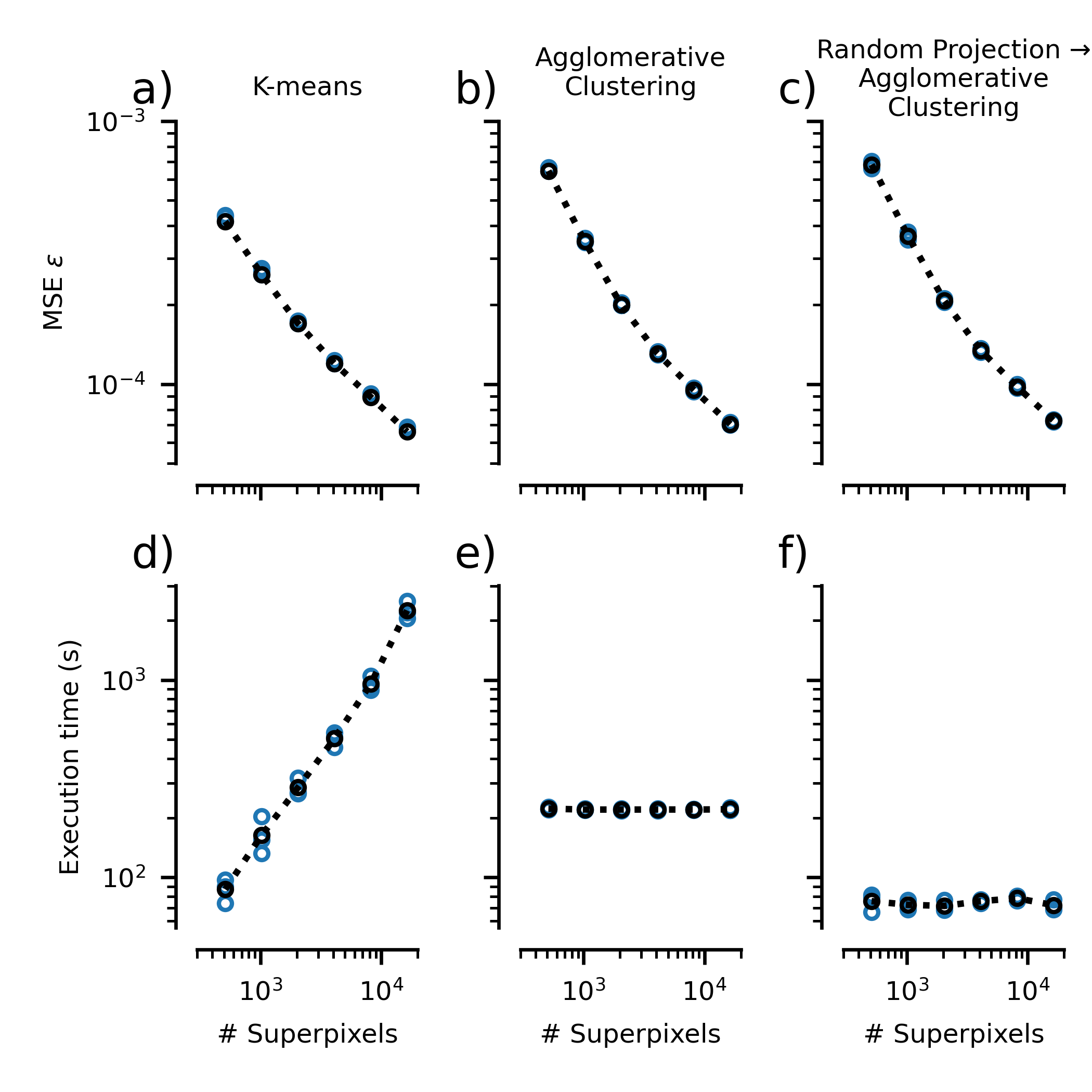}%
\caption{\label{fig_5} Reconstruction error and execution time of different algorithms on large datasets, similar to real-world datasets ($512\times512$ square, 512 features (spatial/temporal measurements) per pixel). Black markers show averages.}
\end{figure}


\subsection{Increasing computational efficiency}
As both K-means and agglomerative clustering are distance-based methods, one can wonder whether it is possible to accelerate the execution time. The Johnson-Lindenstrauss lemma \citep{Johnson1984Extensions} states that it is possible to project a set of points in a high-dimensional space to a lower-dimensional space while \textit{nearly} preserving distances using random orthogonal projections. As we are interested in efficient clustering, and not so much in exact distances, it seems worth investigating whether preprocessing datasets this way results in faster execution. 
512 frames were sampled from the same 4D-STEM dataset. We investigated K-means featurization, agglomerative clustering with connectivity, and gaussian random projection followed by agglomerative clustering with connectivity (where random projection was employed as a preprocessing step) to reduce the dimensionality of the problem (Figure \ref{fig_5}). The execution times reported in Figure \ref{fig_5} encompass the time necessary for random projection (where relevant), clustering of the pixels, and reconstruction of the dataset.
Figure \ref{fig_5} shows that all methods perform similarly in terms of reconstruction error. As mentioned above, more and smaller superpixels provide better sampling of the detector space, resulting in a lower reconstruction error (as also shown directly in Figure \ref{fig_4}). K-means clustering has a lower reconstruction error and execution time when few superpixels are used. However, when employing a sufficient number of superpixels, the connectivity constraints we imposed for the agglomerative clustering increase interpretability without harming the reconstruction error. Furthermore, for K-means clustering, the improved reconstruction error at higher superpixel numbers comes at a significant computational cost, since the amount of operations increases with every iteration. In contrast, for agglomerative clustering the execution time is not impacted by the final amount of superpixels, since the most costly initial agglomeration steps occur in the beginning.

\subsection{Application: preprocessing 4D-STEM data for phase mapping}
\label{sec:nmf}
\label{chiral_ferro}
Parallels can be drawn between this work and our previous work, which used randomized Non-negative Matrix Factorization (RNMF) \citep{erichson2018randomized} to accelerate 4D-STEM phase mapping \citep{Werbrouck2025Fast}. When NMF is used for phase mapping, the `diffraction fingerprint' of each present species is obtained, together with an intensity map for each species, all without prior knowledge or user input other than the amount of components. This makes NMF exquisitely suited for phase mapping in 4D-STEM.

RNMF differs from NMF through a preprocessing step, in which the largest dimension of the data matrix is reduced significantly using QB factorization and Gaussian random projections. Despite the significant acceleration RNMF offers with respect to regular Non-negative Matrix Factorization (NMF), such analyses can be time intensive on larger datasets. 

In this section, we specifically focus on reducing the dimension of the diffraction space in an interpretable way before performing (R)NMF mapping, minimizing the amount of redundant pixels while retaining as much information as possible. For this purpose, we revisit a relatively large 4D-STEM dataset first reported by Ren \textit{et al.} (see Method section) \citep{Ren2025}.

In the original work on this example dataset, it was found that a tight coupling exists between ferroelectric behavior and chiral behavior in \ce{BaTiS3} under applied stress. Real-space pixels were classified into two different orientations using K-means clustering (not related to superpixel construction in diffraction space), excluding the central diffraction spots to minimize the influence of beam intensity variations caused by sample thickness \citep{Ren2025}. 

\begin{figure}[!ht]
\centering
\includegraphics[width=\columnwidth]{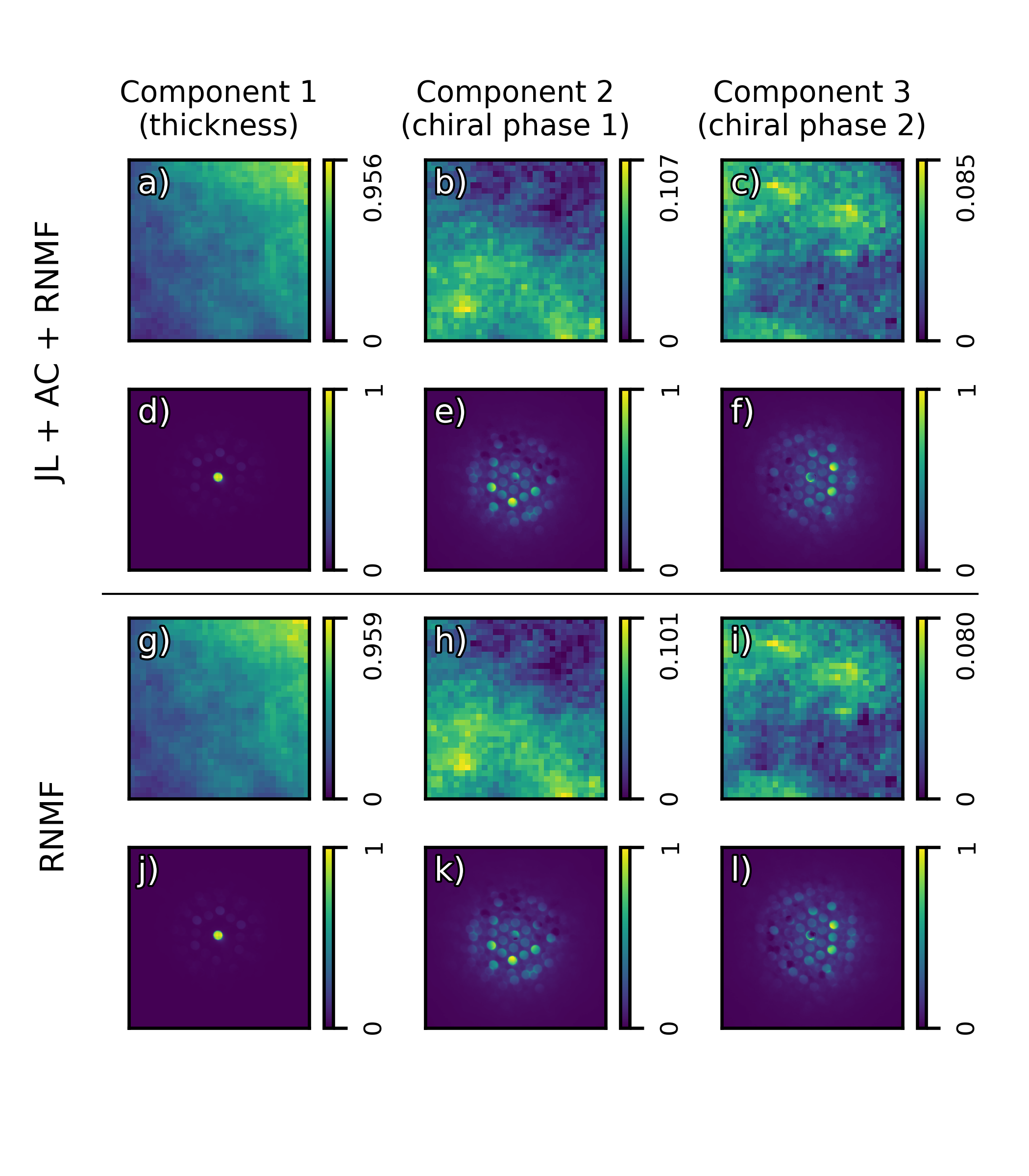}%
\caption{ a-c) Phase maps and d-f) diffraction signatures constructed using RNMF with Johnson-Lindenstrauss/superpixel preprocessing, compared with the g-i) maps and j-l) fingerprints obtained without preprocessing RNMF. Despite the small error from the superpixel construction (\ref{tab_1}), the interpretation is not affected by the superpixel approach. Maps of the reconstruction errors can be found in figure S3.}
\label{fig_6}
\end{figure}

Here, we employ (R)NMF to obtain more detailed maps, showing the spatial distribution and relative concentration of the different phases, while also extracting ferroelectric and orientation information through the more detailed diffraction data. When using three components (Figure \ref{fig_6} a-f), the first component contains thickness information via the direct beam intensity (central spot). The central beam is more intense where the sample is thinner, as a result of reduced scattering. The second and third components then show the diffraction spectrum of the same chiral phase, but rotated $\pi/3$ radians, as expected. We can learn more about the ferroelectric properties of the material by looking at the beam deflections, and indeed, when focusing specifically on the center spot, the second phase shows an upward shift (1 o'clock), while the third phase shows a downward shift (8 o'clock), indicating ferroelectric-induced beam deflection. This is a typical case in which NMF can yield more information than, for example, the K-means approach used earlier. However, this comes at a cost: the wall time for running this analysis was over 3000 s of wall time for 2000 iterations (Table \ref{tab_1}, line 1). Using RNMF, we were able to do 2000 iterations in 150 s. 

To further accelerate this, we grouped the 65536 diffraction pixels into 8192 superpixels through agglomerative clustering (Table \ref{tab_1}, line 2). The supporting information contains a map of the resulting superpixels and their sizes (Figure S2). With this superpixel preprocessing step, the entire analysis now takes about 32.8 s wall time (RNMF), of which 31.1 s are used for superpixel construction. The superpixels generate a small additional contribution to the mean square error. As discussed above, Gaussian random projections can be used before performing pixel clustering to reduce the number of diffraction spectra used for superpixel construction. Implementing Gaussian random projections in this way resulted in a total processsing time of 13.1 s, at the cost of an only slightly higher total reconstruction error for the superpixels (Table \ref{tab_1}, line 3, RNMF). The final phase maps and fingerprints are displayed in Figure \ref{fig_6} g-l. 

We leave the trade-off between computational cost and faithful data representation to the reader. However, we find no difference between interpreting Figure \ref{fig_6} a-f and Figure \ref{fig_6} g-l, whereas the latter is computed more than 20 times faster (based on wall time for RNMF analysis). Compared to regular NMF, the combination of superpixels and QB factorization results in a 250-fold improvement. More importantly, this enables the analysis of larger datasets with more iterations, as discussed in our previous work. 

\begin{table*}[htbp]
\centering
\begin{tabular}{l|cccc|ccc|ccc}
\multirow{2}{*}{Method} & \multicolumn{4}{c|}{Superpixel} & \multicolumn{3}{c|}{RNMF} & \multicolumn{3}{c}{NMF} \\
\cline{2-11}
& \textbf{Pre (s)} & \textbf{AC (s)} & \textbf{Post (s)} & \textbf{MSE} & \textbf{Time (s)} & \textbf{Total (s)} & \textbf{MSE} & \textbf{Time (s)} & \textbf{Total (s)} & \textbf{MSE} \\
\hline
\textbf{As is} & --- & --- & --- & --- & 918.7/150.1& 150.4 & $1.17 \times 10^{-5}$ & 23296.8/3399.2 & 3399.2 & $1.17 \times 10^{-5}$ \\
\textbf{No JL} & --- & 20.0/19.6 & 11.5/11.5 & $5.56 \times 10^{-7}$ & 13.1/1.6 & 32.8 & $1.19 \times 10^{-5}$ & 313.7/39.3 & 70.5 & $1.19 \times 10^{-5}$ \\
\textbf{JL} & 3.6/0.8 & 2.8/2.2 & 8.6/8.6 & $5.87 \times 10^{-7}$ & 10.5/1.3 & 13.1 & $1.19 \times 10^{-5}$ & 290.1/36.4 & 48.1 & $6.53 \times 10^{-4}$ \\
\end{tabular}
\caption{\label{tab_1} Comparison of superpixel construction, RNMF, and NMF computational efficiency. Times shown as CPU/Wall in seconds. The JL column contains the CPU/wall time for the Gaussian random projection (Johnson-Lindenstrauss). The AC column contains the CPU/wall time for the Agglomerative Clustering. The Post column contains the CPU/wall time necessary for constructing the reduced dataset from the superpixels. Total time includes superpixel construction phases (Pre+AC+Post+decomposition) where applicable. The MSE columns aggregate the reconstruction errors associated with the superpixel construction, RNMF and NMF, as also shown in Figure S3. For RNMF, an oversampling of 40 with 4 subspace iterations was used. 2000 iterations were performed and NNDSVDAR initialization was used, both for RNMF and NMF.}
\end{table*}
\section{Discussion}

\subsection{Other algorithms for reciprocal space clustering}
We consider both K-means clustering and agglomerative clustering superior to square binning and/or cropping in terms of information retention. 
It is not surprising to see that they perform very similarly in faithfully reconstructing data, as they solve for the same objective on the same data using top-down and bottom-up strategies, respectively. Due to its bottom-up approach, however, agglomerative clustering was shown to scale better than K-means clustering when many superpixels are employed, and without sacrificing accuracy.

We imagine hybrid approaches here can bring the best of both worlds (i.e. cropping followed by meaningful clustering, as we did for the chiral ferroelectric). Interestingly, azimuthal integration, which is sometimes used as a data reduction technique, can also be thought of as the use of superpixels with an imposed symmetry (when the beamstop is excluded from the integration area). The present work significantly extends this concept beyond azimuthal integration to a general strategy that allows for the retention of 2D information such as Bragg peaks.

As both K-means and agglomerative clustering strategies exploit Euclidean distances, we demonstrate that they can be further accelerated using Gaussian random orthogonal projection to reduce the feature space prior to clustering. Other approaches, such as random sampling, are possible: Clustering on a random subset of data, followed by batched processing on the rest of the data could be an interesting way to process extremely large datasets, providing an acceleration in superpixel identification compared with using the full dataset. 

It could be possible to further accelerate this analysis by using GPUs, or by using other clustering algorithms to optimize the same objective. We note that not every clustering method is suitable for our goals. Single-linkage clustering methods such as DBSCAN \citep{Ester1996Density}, or single-linkage agglomerative clustering are significantly faster, but do not minimize the intra-cluster variance and are thus less suitable for faithful reconstruction purposes (Figure S4). 

\subsection{Advantages and disadvantages of superpixels}
Superpixels with connected agglomerative clustering provide a drop-in, interpretable and scalable preprocessing step for existing pipelines. Even though superpixels do not provide interpretation by themselves,  they indicate regions (hence, provide features) with similar information density. In contrast to Bragg disc featurization, superpixels retain the information  in `empty reciprocal space' as well. 

Additionally, we note that it is easily possible to construct a \textit{graph} from a featurization with superpixels. A square grid forms a trivial graph with $n$ nodes and $4n$ edges (which was exploited in creating the superpixels with connectivity). In contrast, a set of superpixels can be cast as a much smaller graph with $k$ nodes and $\ll 4n$ edges. This opens the door to data interpretation with graph neural networks, where reciprocal superpixels are not treated as standalone features but internal relations between the features are preserved. 

Despite its advantages, the construction of superpixels does involve an additional processing step, with associated loss of information. A suitable parameter $k$ also must  be determined by the user. Finally, there are some potential reasons that could weaken our core assumptions, such as beam movement, vibrations, and electromagnetic coil hysteresis. We believe that these non-idealities have similar effects on other processing approaches. 

\section{Conclusions}

4D-STEM or synchrotron experiments can yield (very) large datasets (1-100 GB) from a single measurement of a material sample. These datasets typically contain hidden symmetries, rendering the information density per pixel rather low. Additionally, high-information density areas and low-information density areas are persistent across entire diffraction datasets. We identify that these properties set apart diffraction datasets (including 4D-STEM, in situ, or operando diffraction experiments) among other large datasets consisting of a series of 2D data (such as video). 

We exploit the symmetries in diffraction data by grouping and averaging pixels into superpixels, using methods based on minimizing intra-cluster variance. This effectively reduces the size of the problem by a factor defined by the superpixel-pixel ratio. The scaling behavior and interpretability make agglomerative clustering preferable over K-means clustering for diffraction data. When the local structure is taken into account, it is as if each dataset was measured with a custom, tailored detector for target features relevant to the sample, offering high sampling densities (small superpixels) in regions with high information, and lower sampling densities (large pixels) elsewhere. 

Developing this workflow culminated in the large acceleration (100-to 200-fold) of the analysis of the \ce{BaTiS3} dataset, whereas the RNMF methodology provided a more detailed insight in chiral-ferroelectric coupling, corroborating and extending the original interpretation of the results.

The simple interpretation and unsupervised training approach make superpixels well suited for the goal of preprocessing large diffraction datasets into datasets of significantly smaller size while preserving information. These demonstrations pave the way for further adoption of next-generation AI/ML-based analyses on diffraction datasets. 


\section{Competing interests}
No competing interest is declared.

\section{Author contributions statement}
A.W. and M.J.Y. conceived the project. A.W. developed the code and conducted the analysis. N.C.P. prepared the samples. X.H. collected the microscopy data. A.M. assisted with the analysis of the ferroelectric data. A.W. and M.J.Y. wrote and reviewed the manuscript. 

\section{Acknowledgments}
The 4D-STEM \ce{BaTiS3} dataset was kindly provided by Guodong Ren. A.W. acknowledges support from the University of Missouri Materials Science and Engineering Postdoctoral Fellowship. N.C.P. and M.J.Y. acknowledge partial support from the National Science Foundation Electrochemical Systems Program through Award Number 2219060. M.J.Y. acknowledges partial support from the National Science Foundation Division of Graduate Education NRT Program through Award Number 2243526. 

A preprint version of this manuscript was previously posted at \url{https://doi.org/10.48550/arXiv.2607.13979} \citep{werbrouck2026using}.

\bibliographystyle{plainnat}
\bibliography{lib}
\end{document}